\documentclass{svjour3}                     
\smartqed  
\usepackage{graphicx}
 \journalname{Scientometrics}
\begin{document}

\title{Editorial process in scientific journals: \\analysis and modeling
}

\titlerunning{Editorial process in scientific journals}        

\author{O. Mryglod   \and Yu. Holovatch \and I. Mryglod
}


\institute{O. Mryglod \at
              Institute for Condensed Matter Physics of the NAS of Ukraine \\
              1 Svientsitskii Str., 79011 Lviv, Ukraine\\
              \email{olesya@icmp.lviv.ua}
           \and
            Yu. Holovatch \at
              Institute for Condensed Matter Physics of the NAS of Ukraine \\
              1 Svientsitskii Str., 79011 Lviv, Ukraine
           \and
           I. Mryglod \at
              Institute for Condensed Matter Physics of the NAS of Ukraine \\
              1 Svientsitskii Str., 79011 Lviv, Ukraine
}

\date{Received: date / Accepted: date}

\maketitle

\begin{abstract}
The editorial handling of papers in scientific journals as a human
activity process is considered. Using recently proposed approaches
of human dynamics theory we examine the probability distributions
of random variables reflecting the temporal characteristics of
studied processes. The first part of this paper contains our
results of analysis of the real data about papers published in
scientific journals. The second part is devoted to modeling of
time-series connected with editorial work. The purpose of our work
is to present new object that can be studied in terms of human
dynamics theory and to corroborate the scientometrical application
of the results obtained.

\end{abstract}

\section*{Introduction}
\label{intro} Nowadays new possibilities for quantitative studying
of human activity processes appear. Modern computer technologies
allow to collect and to process large volumes of statistical data.
It is possible to fix the time of each performed operation from
large flow of human (customer/user/client) actions and to obtain
the data set representing whole picture. Therefore, new methods
for data analysis and knowledge discovering could be applied. In
particular, this allows quantitative analysis of psychological,
social and other aspects of human behavior.

The sequences of human actions (telephone calls, information
queries or stock exchanges) are not new subject to study. But
contrary to generally accepted opinion about human actions
randomly distributed in time and thus well approximated by Poisson
processes (interevent times are exponentially distributed), the
comparatively recent results of human dynamics analysis show the
presence of power laws which nature and origin are unexplained up
till now
\cite{2005_Barabasi_Nature,2005_Oliveira&Barabasi_Nature,2008_Zhou}.
This discovery attracts an interest and provokes analysis of new
processes involving human actions in different fields of our life.
The analysis of time statistics of human activity patterns can be
useful for different optimization and control tasks in spheres of
mass service, communication, information technologies, resource
distribution, etc. Besides, this approach can give a new
possibility to understand human behaviour and to get its
additional quantitative measure.

The continuous human behaviour can be considered as the set of
consecutive actions in time. In this case it is convenient to
examine two characteristic random variables: the time interval
between two consecutive actions (called the interevent time
$t_{\mathrm{int}}$) and the time a task is waiting for an execution
(the so-called waiting time $t_{\mathrm{w}}$). The power-law nature
of corresponding distributions
\begin{equation}
P(t_{\mathrm{int}})\sim t_{\mathrm{int}}^{-\alpha}, \qquad
P(t_{\mathrm{w}})\sim t_{\mathrm{w}}^{-\beta}, \qquad
\alpha, \beta >0
\label{power-law}
\end{equation}
arises from the analysis of real processes that involve human
actions such as browsing the Internet, data downloading, electronic
and mail communication, initiating financial transactions etc.
\cite{2005_Barabasi_Nature,2005_Oliveira&Barabasi_Nature,2004_Johansen}.
On the other hand, the processes with execution of tasks in queue
are the typical objects of study in the queueing theory which can
be seen as the subfield of the so-called theory of mass service
or, generally speaking, of applied probability theory
\cite{2006_Vazquez,1981_Cooper}. So, it is natural to use some
general terminology to describe such kinds of processes.
An empirical analysis results in a variety of values of the
exponents governing (\ref{power-law})
\cite{2008_Zhou,2006_Vazquez}, however an observed tendency when
similar values of the exponents describe different processes gives
rise to an analogy with the ``universality classes'' found in
physics of critical phenomena \cite{2006_Vazquez}. Different
models have been proposed to gain insight about an origin of power
laws (\ref{power-law}) \cite{2008_Zhou}.
Some of the proposed models are based on the assumption about the
key role of priority-based (decision-based) queuing process
\cite{2006_Vazquez}. The situation when an individual has to
perform a list of tasks and chooses a task from this list using
some internal priority is very natural. It is important, that
power-law-like distributions (\ref{power-law}) appear only in the
so-called critical and supercritical regimes of service, when its
traffic intensity $\rho$:
\begin{equation}
\rho=\lambda/\mu \label{traffic_int}
\end{equation}
is greater than 1. In (\ref{traffic_int}) $\lambda$ is the tasks
arrival rate and $\mu$ is the execution rate, respectively
\cite{2006_Vazquez}. In other words, the probability distribution
functions of human activity processes are close to power law only
when the queue of tasks is not exhausted \cite{2006_Vazquez}. This
class of human activity can be described by the so-called
``task-driven'' models \cite{2008_Zhou}. On the other hand, the
``interest-driven'' models explain the existence of power laws in
human dynamics
 in a different way. There, the power laws may
be observed also for the processes without any possibility to
determine the queue of tasks, for example visiting of web-sites by
different users \cite{2008_Zhou}.

Thus, we can conclude that human dynamics theory is the issue of
current importance. Often besides academic interest there is an
obvious great practical value of the results obtained.

The main purpose of this paper is to gain insight on the editorial
process in scientific journals using tools and concepts of the
human dynamics theory. In the first part
(section~\ref{Experiment}) we show our empirical results obtained
during the data analysis about distributions of waiting times of
papers in several scientific journals. The simple simulation model
of editorial processing of scientific manuscripts is presented in
the second part of this paper (section~\ref{Model}).

\section{Waiting times statistics analysis of scientific journals}
\label{Experiment} 
\begin{figure}[ht]
\centerline{\includegraphics[width=10cm]{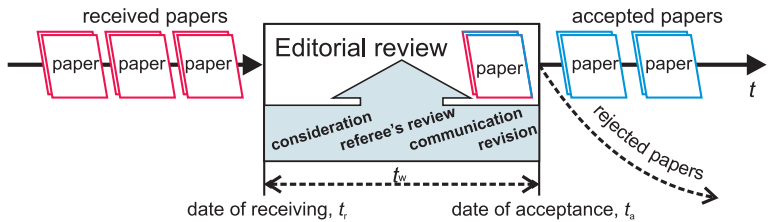}}
\caption{Schematic picture of editorial processing of manuscripts
in scholarly journals.} \label{Fig1}
\end{figure}

Let us consider the editorial process in
scientific journals as an example of human activity processes
\cite{2007_Mryglod}. Keeping in mind the classification mentioned
above, we will consider the ``task-driven'' model. In this kind of
mass service system the input flow consists of submitted papers
forming the queue. A standard procedure after paper submission can
include the following steps: (i) peer-review process, (ii)
revisions if necessary, (iii) acceptance by an Editorial Board,
(iv) other intermediate processes. On each of the above stages the
paper may be rejected. However, typically the information
concerning the rejected papers is not publicly available.
Therefore, we consider the random variable $t_{\mathrm{w}}$
defined as a number of days between the dates of the paper final
acceptance $t_{\mathrm{a}}$ and the paper receiving
$t_{\mathrm{r}}$:
\[
t_{\mathrm{w}}=t_{\mathrm{a}}-t_{\mathrm{r}}.
\]
All stages of the editorial processing of submitted manuscripts
are considered together as the one process (Fig.~\ref{Fig1}).
Though more than one actor takes part in it, we consider that
every part of this work is controlled by an Editorial Board.
Therefore, we can treat this process as one of the main
characteristics of the Editorial Board activity.

During the real data analysis we met the problem of acceptance
time $t_{\mathrm{a}}$ determining whereas the dates of paper
receiving $t_{\mathrm{r}}$ are usually fixed very accurately.
Different dates reflecting stages of papers processing are
available for different scientific journals: date of revision,
date of final acceptance, date of availability on-line, etc. It is
necessary to specify the meaning of $t_{\mathrm{a}}$ for every
particular journal. Besides, it is also interesting to consider
various sets of $t_{\mathrm{w}}$ choosing different meanings of
$t_{\mathrm{a}}$ for the same journal and to compare the obtained
results. In our analysis, the $t_{\mathrm{a}}$ was defined as the
date of revised version if the final acceptance date was omitted
and as the final acceptance date (if present). So,
$t_{\mathrm{w}}$ were calculated as the time differences between
the date of submission and the most distant of two possible dates:
revision date or final acceptance date.

Our goal was to determine the functional form of probability
distributions $P\left(t_{\mathrm{w}}\right)$ based on the
statistical data analysis performed for a few scientific journals.
Another task was to find if possible a typical form of
$P\left(t_{\mathrm{w}}\right)$ for normally working Editorial
Board.

\begin{figure}[ht]
\centerline{\includegraphics[width=8cm]{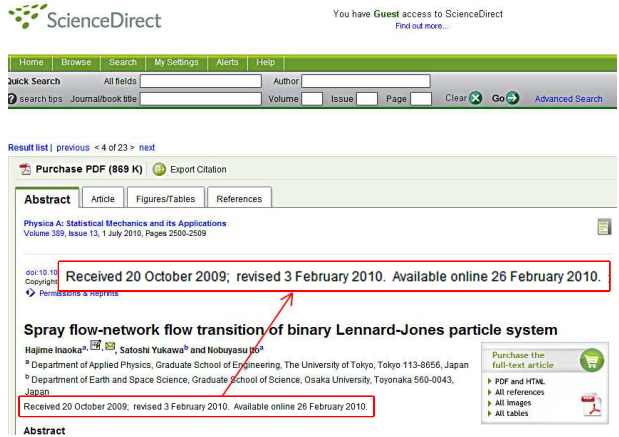}}
\caption{An example of publicly available data (the official
web-site of the journal ``Physica~A'') which were used for our
research.} \label{Fig2}
\end{figure}

Several journals with different Editorial Boards were chosen in
our study: three of them belong to the international Elsevier
Publishing House (``Physica~A: Statistical Mechanics and its
Applications'', ``Physica~B: Condensed Matter'' and ``Information
Systems'') \cite{2007_Mryglod}. One more journal is rather new
``Condensed Matter Physics'' (published by the Institute for
Condensed Matter Physics: www.icmp.lviv.ua).
 The publicly available data from the
official web-sites were used for analysis (Fig.~\ref{Fig2}). Also
we tried to get the corresponding statistics of
``Scientometrics'', although the data set is too small because the
required dates for papers are available online only for several
last years.

Some formal parameters of databases used in this part of our work
are shown in table~\ref{tab1_DB_1}. As we can see from the table,
the typical number of  records for each journal is of order of
$10^3$ which allows to make some quantitative conclusions. These
general conclusions hold even for a more poor statistics (c.f.
results for ``Condensed Matter Physics'' in Fig.~\ref{Fig6_CMP}).
Zeros in the table mean that for some papers their submission date
is coincide with the final acceptance date. But what is even more
interesting, in every journal one can find papers with very long
waiting periods (for example, 2260 days it is more that 6 years!).
Oh course, we can only speculate about the possible reasons for
that. For example, it could be long discussion about the
manuscript or banal misprint.
\begin{table}[ht]
\caption{Characteristics of the data sets analysed: general number
of records, maximal ($t_{\mathrm{w}}^{\mathrm{max}}$), minimal
($t_{\mathrm{w}}^{\mathrm{min}}$), typical
($t_{\mathrm{w}}^{\mathrm{\mathrm{typ}}}$), mean
($t_{\mathrm{w}}^{\mathrm{\mathrm{mean}}}$), and median
($t_{\mathrm{w}}^{\mathrm{\mathrm{med}}}$) waiting times for the
journals under consideration.}
\label{tab1_DB_1}
\begin{center}
\begin{tabular}{|c|c|c|c|}
\hline
&``Physica A''&``Physica B''&``Information Systems''\\
&(1975--2010)&(1988--2010)&(1975--2010)\\
\hline Number of records&4576&4944&814\\
\hline
$t_{\mathrm{w}}^{\mathrm{max}}$, days&1629$^{*}$&1087$^{\ddag}$&2260$^{\dag}$\\
\hline
$t_{\mathrm{w}}^{\mathrm{min}}$, days&0&0&0\\
\hline
$t_{\mathrm{w}}^{\mathrm{\mathrm{typ}}}$, days&60&80&245\\
\hline
$t_{\mathrm{w}}^{\mathrm{\mathrm{mean}}}$, days&124&122.2&331.7\\
\hline
$t_{\mathrm{w}}^{\mathrm{\mathrm{med}}}$, days&95&90&275\\
\hline
\end{tabular}
\end{center}
$^{*}$ I.A.~McLure, A.-M.~Williamson, Physica A, 1996,
\textbf{234}, Iss.~1--2, 206--224; ibid. 225--238.\\
$^{\ddag}$ V.G.~Bar'yakhtar, V.A.~Popov, Physica B, 1999,
\textbf{269}, Iss.~2, 123--138.\\
$^{\dag}$ M.~Binbasioglu, D.~Karagiannis, Information Systems,
2000, \textbf{25}, Iss.~6--7, 453--463.
\end{table}

 At the first stage the probability histograms of waiting times
$P(t_{\mathrm{w}})$ for selected journals were constructed
\cite{2007_Mryglod}. All experimental values were distributed
among discrete intervals (bins) of length equals to 5 days.

\begin{figure}[ht]
\begin{center}
\includegraphics[width=0.485\textwidth]{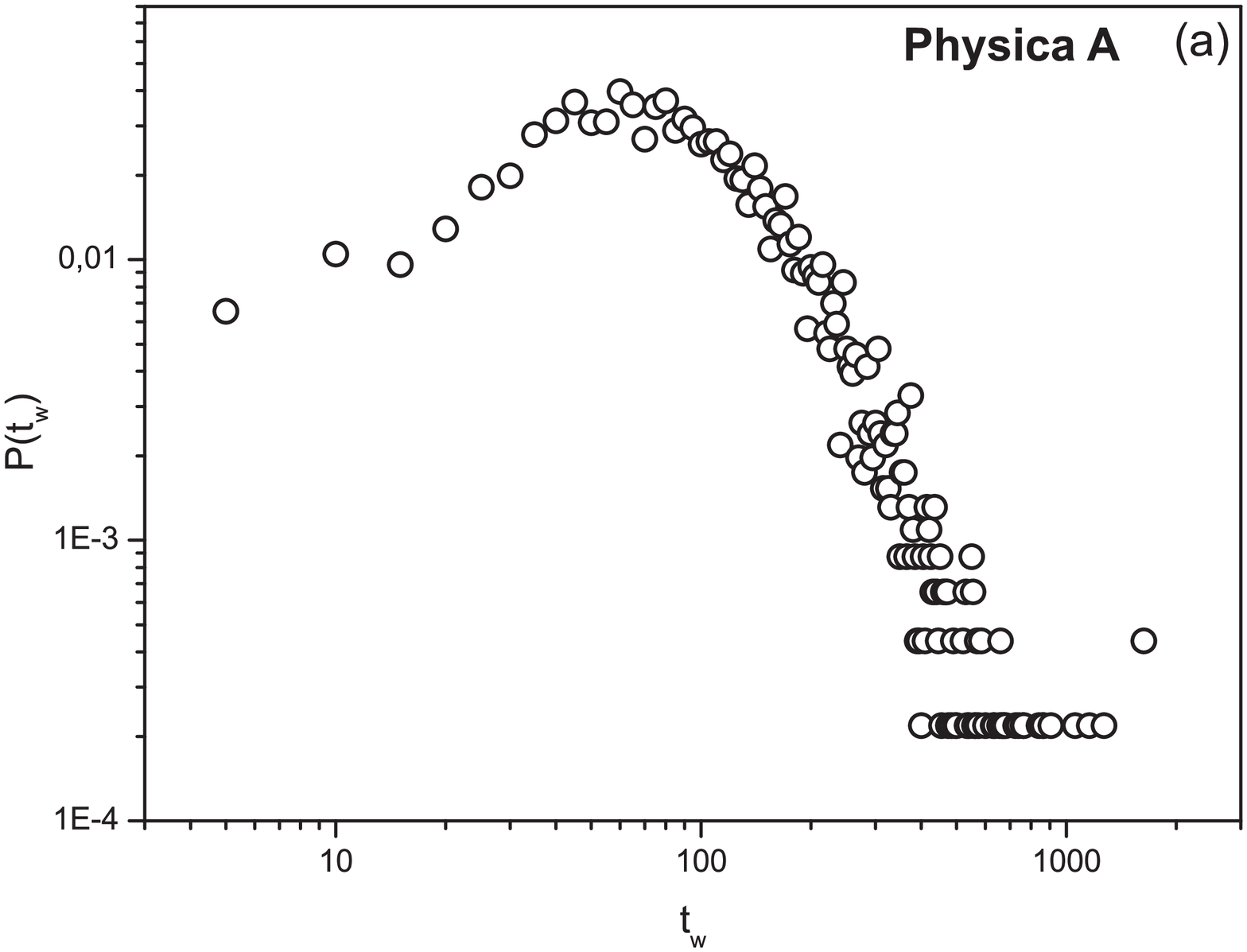}%
\hfill%
\includegraphics[width=0.475\textwidth]{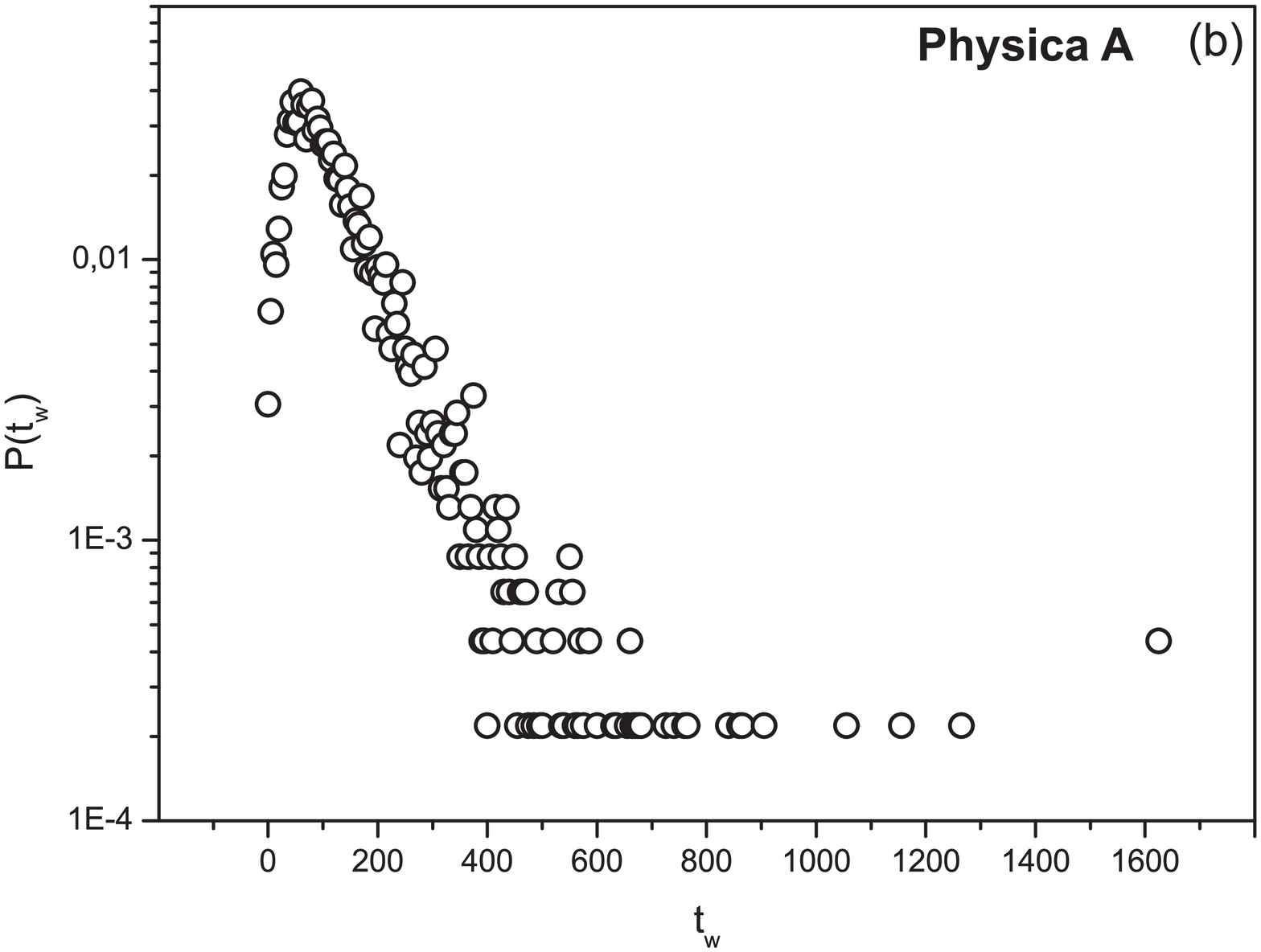}
\end{center}
\caption{(a) Log-log plot and (b) linear-log plot of the
$P(t_{\mathrm{w}})$ distribution  for journal ``Physica~A''.}
\label{Fig3_PhA}
\end{figure}

 Our purpose was to verify the
functional form of $P(t_{\mathrm{w}})$ and to refer it to the
power-law-like class (non-Poisson processes) or, for example, to
the exponential-like class (Poisson processes). The exponential
distributions of random variables $t_{\mathrm{int}}$ and
$t_{\mathrm{w}}$ are evidences of the random selection of tasks to
execute \cite{2008_Zhou,2005_Barabasi}. In Fig.~\ref{Fig3_PhA} we see 
the $P(t_{\mathrm{w}})$ distribution  for journal ``Physica~A''.
It has a form of a unimodal non-symmetrical distribution with a
smooth decay. Since the distribution is skew, the mean value
$t_{\mathrm{w}}^{\mathrm{mean}}$ does not coincide with the
typical value $t_{\mathrm{w}}^{\mathrm{typ}}$, at which maximum in
$P(t_{\mathrm{w}})$ occurs. Both the mean, typical and median
values of $t_{\mathrm{w}}$ are given in the table~\ref{tab1_DB_1}.
The log-log and log-linear plots of Fig.~\ref{Fig3_PhA}
demonstrate good possibilities for linear approximations of
$P(t_{\mathrm{w}})$ in both scales and this situation is analogous
for other journals analysed \cite{2007_Mryglod}.

\begin{figure}[ht]
\begin{center}
\includegraphics[width=0.485\textwidth]{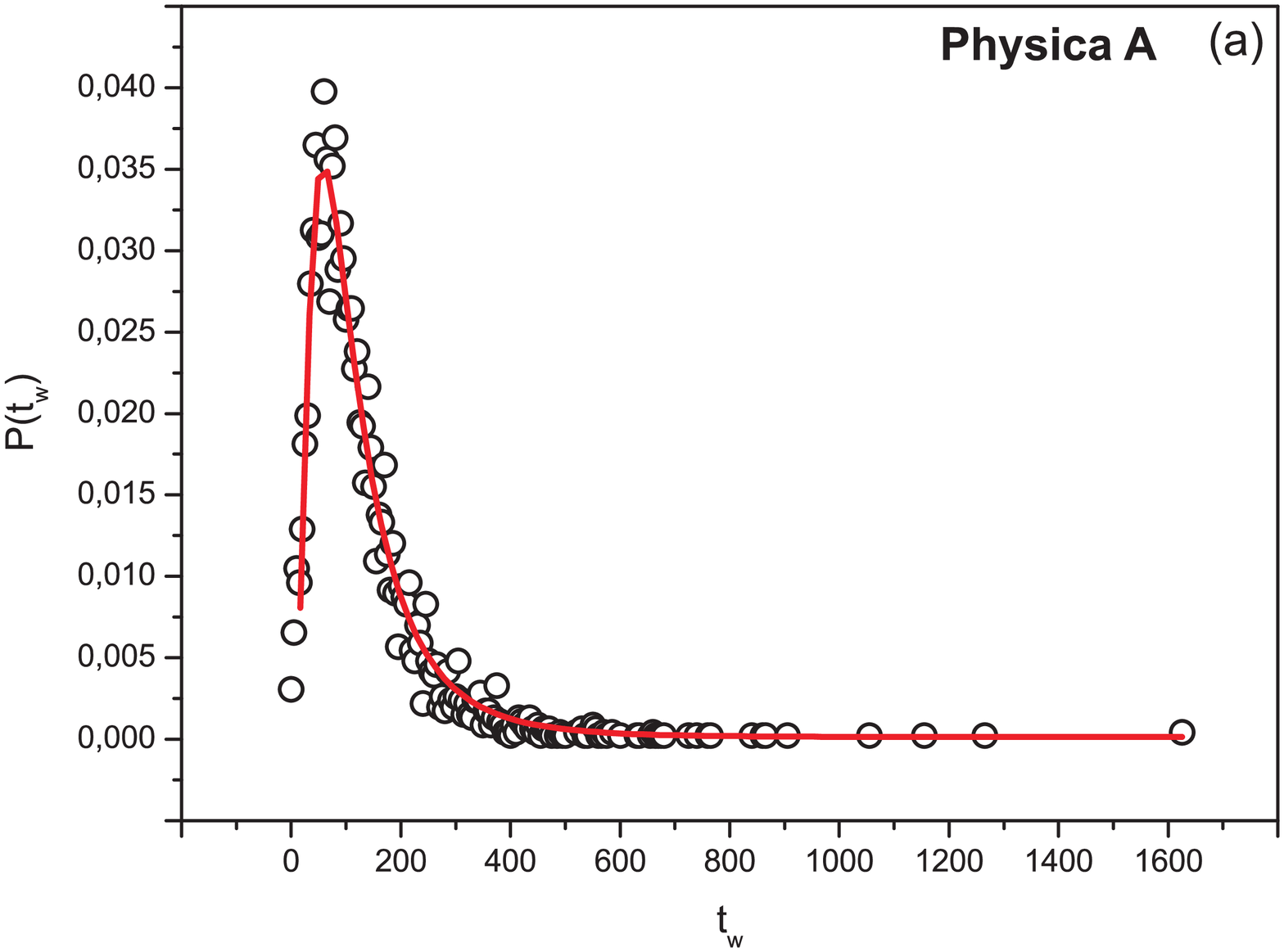}%
\hfill%
\includegraphics[width=0.475\textwidth]{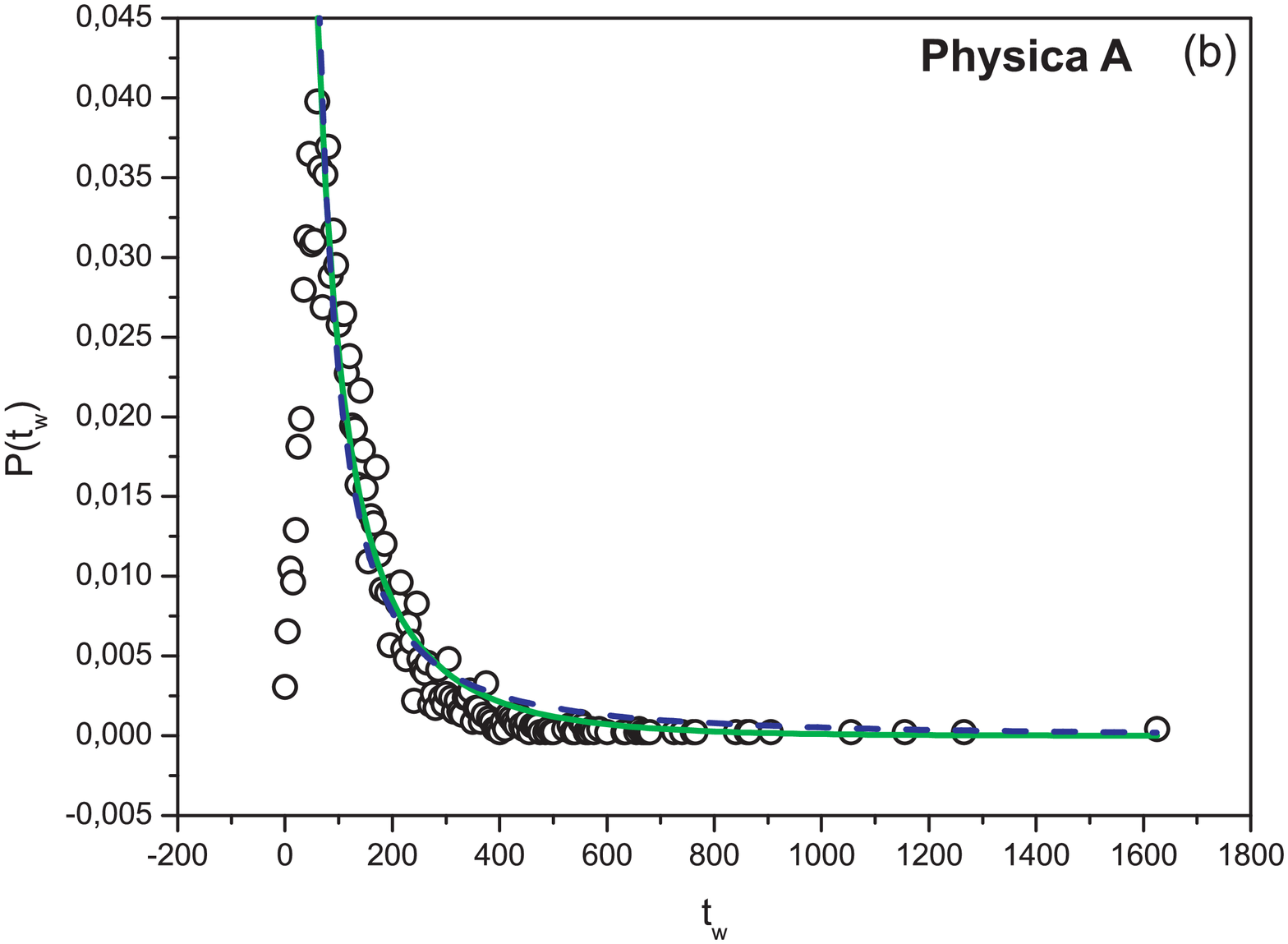}
\end{center}
\caption{The $P(t_{\mathrm{w}})$ distribution for journal
``Physica~A'' with different approximation curves: (a)
log-normal~(\ref{log-normal}) (solid line, red online), (b)
power-law with exponential cutoff~(\ref{power-cut-off}) with
$\alpha=1$ (light solid line, green online) and $\alpha=3/2$
(dashed line).} \label{Fig4_PhA}
\end{figure}

Further, we have verified two main hypotheses about the form of
probability distributions which are used to describe human
activity processes: (i) log-normal
distribution~\cite{2006_Stouffer}:
\begin{equation}
P(t_{\mathrm{w}})=P_0+\frac{A}{\sqrt{2\pi}\omega
t_{\mathrm{w}}}\mathrm{e}^{-\frac{\left[\ln{\left(\frac{t_{\mathrm{w}}}{t_{\mathrm{c}}}\right)}\right]^2}{2\omega^2}},
\qquad
t_{\mathrm{c}}, \omega >0, \label{log-normal}
\end{equation}
where $\ln{(t_{\mathrm{c}})}$ and $\omega$ are the mean and standard
deviations of the $\ln{(t_{\mathrm{w}})}$, $P_0$, $A$ are fitting
constants; %
and (ii) power-law distribution with exponential cutoff
\cite{2006_Vazquez} for exponent values $\alpha=\{1; 3/2\}$:
\begin{equation}
P(t_{\mathrm{w}})=At_{\mathrm{w}}^{-\alpha}\mathrm{e}^{-\frac{t_{\mathrm{w}}}{t_0}},
\qquad%
t_0>0, \label{power-cut-off}
\end{equation}
where $t_0$ is characteristic of waiting time which depends on
traffic intensity, $A$ is a constant. Using fitting procedure we
found optimal parameters for both distributions (\ref{log-normal})
and (\ref{power-cut-off}). The results of fits obtained for
journal ``Physica~A'' are shown in Fig.~\ref{Fig4_PhA} by smooth
curves.
\begin{figure}[!h]
\begin{center}
\includegraphics[width=0.485\textwidth]{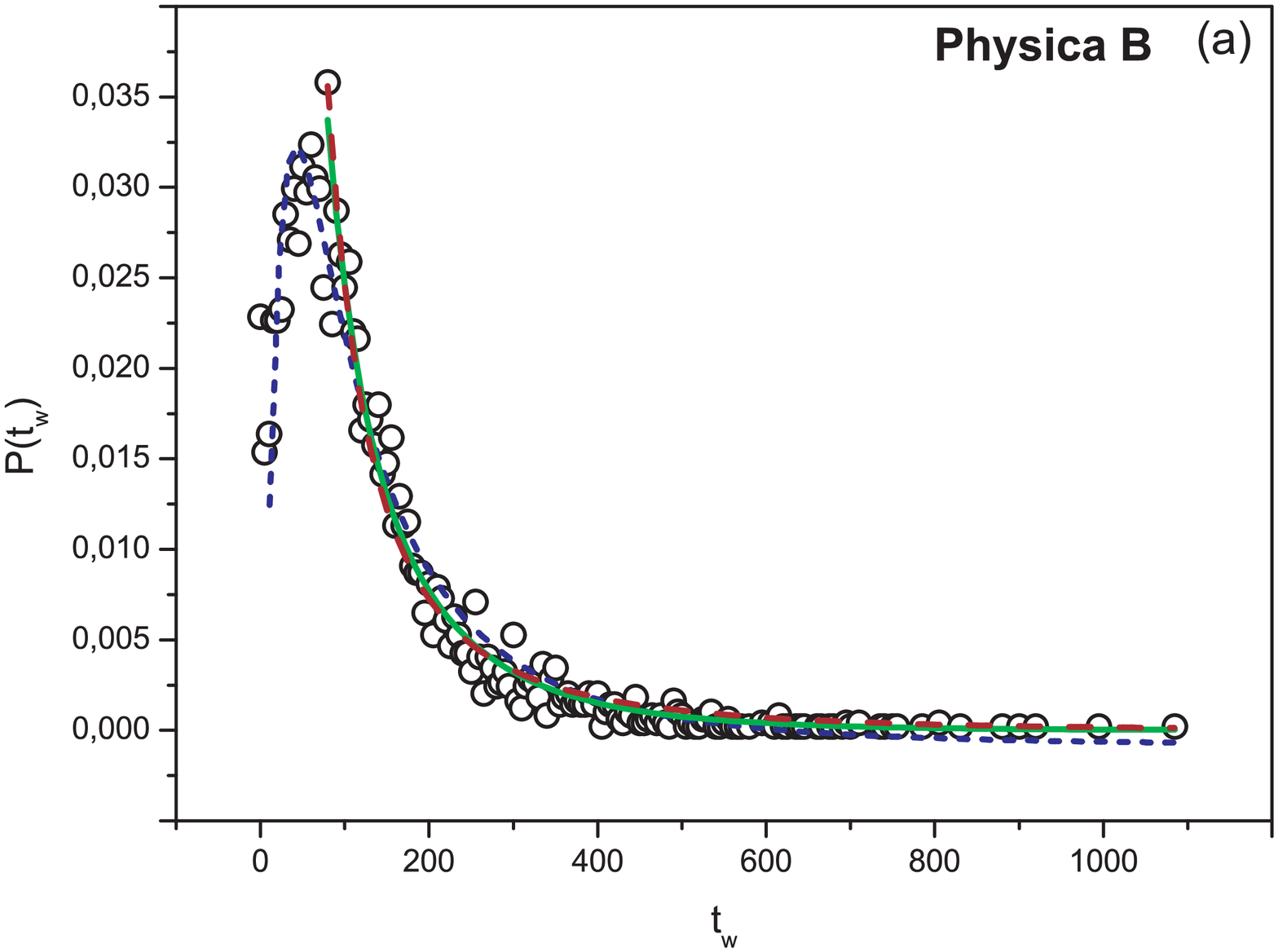}%
\hfill%
\includegraphics[width=0.475\textwidth]{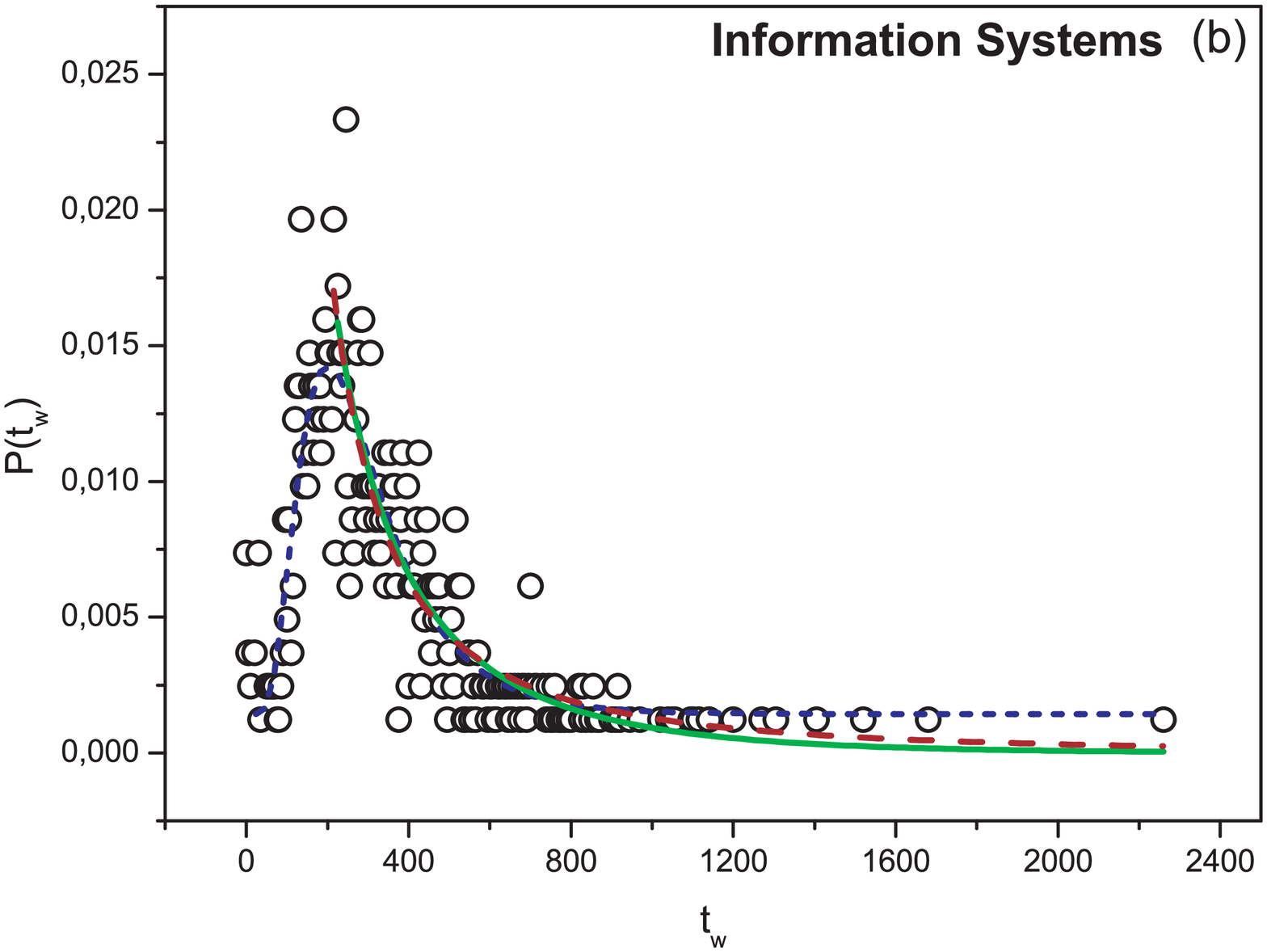}
\end{center}
\caption{The $P(t_{\mathrm{w}})$ distribution for (a)
``Physica~B'' and (b) ``Information Systems'' journals. The dotted
dark (blue online) lines  are the approximations by
log-normal~(\ref{log-normal}), the light solid (green online) and
dark dashed (red online) lines  -- by power-law with exponential
cutoff~(\ref{power-cut-off}) with $\alpha=1$ and $\alpha=3/2$,
respectively. }
\label{Fig5_PhB_IS}
\end{figure}

To compare the accuracy of approximations by different functions
(log-normal~(\ref{log-normal}) and power-law with exponential
cutoff (\ref{power-cut-off})) we exploited the criterium based on
the value of the adjusted coefficient of determination $\bar{R}^2$
\cite{R_sqr}. This coefficient is used to verify the closeness of
experimental data to the non-linear theoretical curve and is a
modification of statistical coefficient of determination ${R}^2$
which is the square of the sample correlation coefficient between
the outcomes and their predicted values.
A value of $\bar{R}^2$ close to 1 indicates that the fit is a good
one. For example, for the journal ``Physica~A'' we have found that
both log-normal (\ref{log-normal}) ($\bar{R}^2\approx 0,97$) and
power-law function with exponential cutoff (\ref{power-cut-off})
and exponent $\alpha=1$ ($\bar{R}^2\approx 0,95$) can be the
probable functions of distributions $P(t_{\mathrm{w}})$. The value
of $\bar{R}^2$ for power-law approximation function with
exponential cutoff (\ref{power-cut-off}) and exponent $\alpha=3/2$
is slightly smaller ($\approx 0,92$) for this journal.

In fact, both log-normal and power-law functions predict the same
leading behavior $t^{-1}$, differing only in the functional form
of the exponential correction \cite{2005_Barabasi}. The advantage
of hypothesis about log-normal functional form of
$P(t_{\mathrm{w}})$ consist in the possibility to describe all the
data span, not only the tail.
\begin{figure}[!h]
\begin{center}
\includegraphics[width=0.485\textwidth]{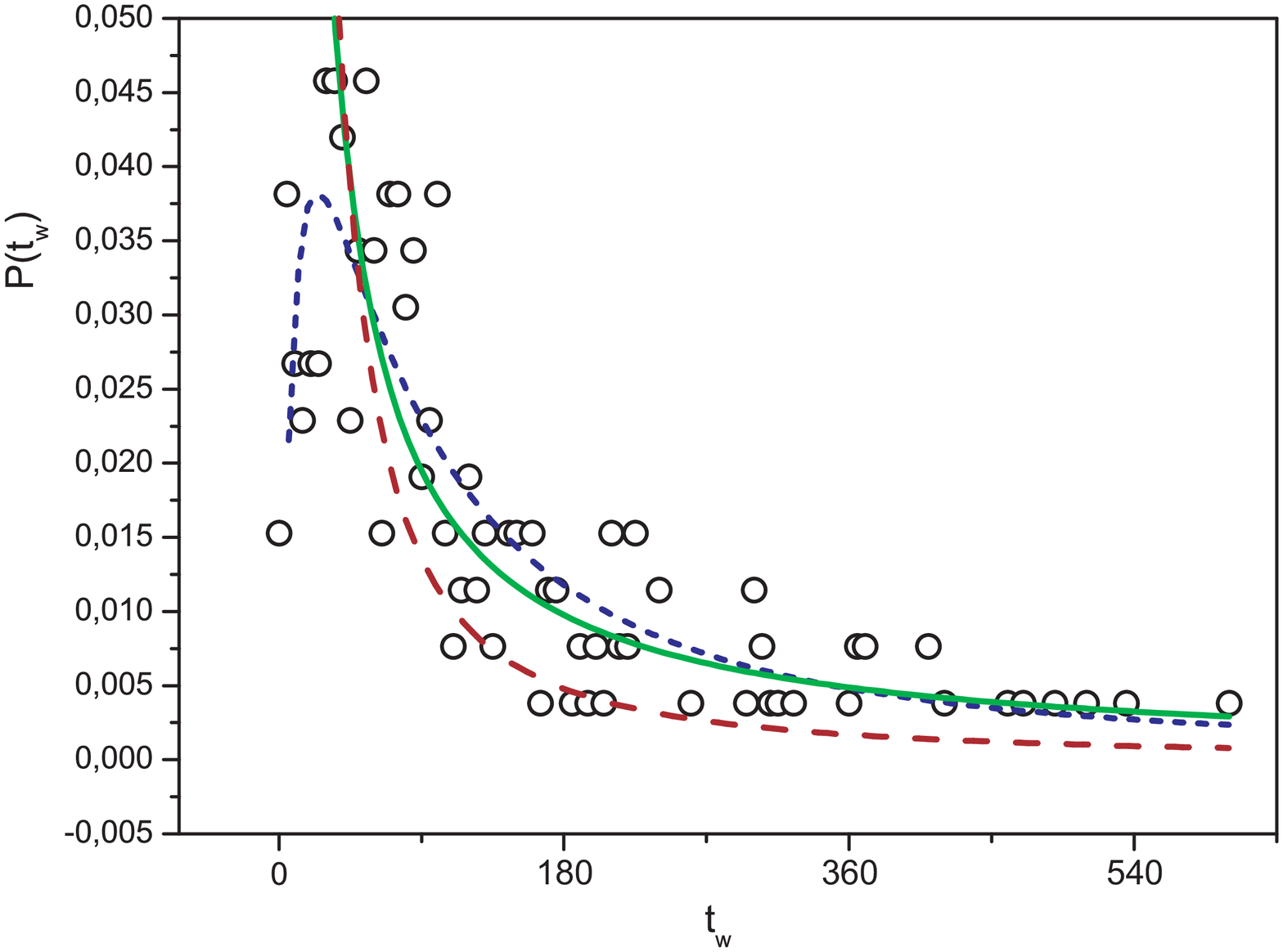}%
\end{center}
\caption{The $P(t_{\mathrm{w}})$ distribution for ``Condensed
Matter Physics'' journal. The dotted dark lines (blue online) are
the approximations by log-normal~(\ref{log-normal}), the light
solid (green online) and dark dashed (red online) lines -- by
power-law with exponential cutoff~(\ref{power-cut-off}) with
$\alpha=1$ and $\alpha=3/2$, respectively. }
\label{Fig6_CMP}
\end{figure}
The results of analogous approximations obtained for ``Physica~B''
and ``Information Systems'' journals are presented in
Fig.~\ref{Fig5_PhB_IS}. For all journals analysed the conclusions
are common: the both hypothesis about possible data approximations
by log-normal~(\ref{log-normal}) and by power-law with exponential
cutoff (\ref{power-cut-off}) and $\alpha=1$ are almost equally
good. Moreover, these conclusions are also right for the
``Condensed Matter Physics'' journal in spite of much smaller
database with only 262 records (see Fig.~\ref{Fig6_CMP}). The
observed data fluctuations can be explained by relatively small
statistics but such situation is usual for the majority of
scientific journals.

We could resume that it is hard to discriminate between the
power-law or exponential nature of the tail of the
$P(t_{\mathrm{w}})$ distributions. But as is easy to see from
obtained results the functional form of these distributions is the
same: one explicit maximum and the long tail with several large
values of $t_{\mathrm{w}}$. The tail of the distributions built
for several journals could be well approximated by functions with
an exponential cut-off and leading power-law behavior $t^{-1}$.
Thus, we could consider the obtained form of probability
distributions $P(t_{\mathrm{w}})$ as the typical one that can be
used for scientometrical analysis of a given journals.

\section{Editorial process modeling}
\label{Model} We came to conclusion about the typical form of
waiting time distributions $P(t_{\mathrm{w}})$ for scientific
journals with normally working Editorial Boards based on the
results described above.
%
%
The origin of such functional form of $P(t_{\mathrm{w}})$
distributions is unknown. We can suppose that the contribution of
human dynamics could be the reason of observed affinity of
$P(t_{\mathrm{w}})$ to power-laws. The peer-reviewing stage
(including the work of authors and the communication process) is
the human activity probably most similar to ``natural'' in
comparison with the rest stages of editorial process. The other
phases consist of different periodical tasks (Editorial Board
meetings, uploads to web-site, etc.) or work with manuscripts in
order of receiving (i.e., language and technical editing).
\begin{figure}[!h]
\begin{center}
\includegraphics[width=0.98\textwidth]{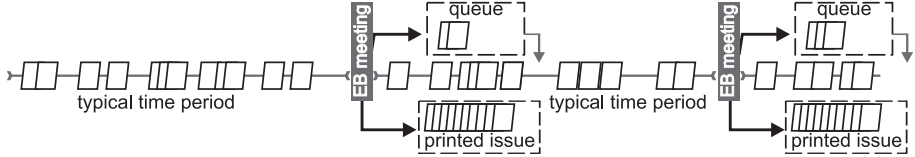}%
\end{center}
\caption{The schematic representation of modelled editorial work in scientific journals.}
\label{Fig7_process}
\end{figure}

To verify the hypothesis about the key role of peer-reviewing in
the waiting times distributions $P(t_{\mathrm{w}})$ shaping we
built the simple simulation model of editorial work in scientific
journal omitting the peer-reviewing \cite{2008_Mryglod}. Below we
show the (expected) crucial difference of the waiting time
distributions in such a model from the real data set analysed in
the previous section. In the frame of this model, we consider the
input flow of the submitted papers. Decision about manuscript
acceptance is taken during regular meetings of the Editorial
Board. All papers submitted before the meeting are considered
(i.e. accepted or rejected) during this meeting. Generally, the
Editorial Board meetings are held periodically and at least once
before publishing of each issue.
Let us defined the typical journal period $T$ as a time interval
(in days) between two consecutive journal issues.
For example, since 1988 the journal ``Physica~A'' has 24 issues
per year and in this case $T=15$~days. For ``Condensed Matter
Physics'' journal (Fig.~\ref{Fig6_CMP}) $T=90$~days since 1997
year.
So, in our model the period between two consecutive meetings
equals to the typical journal period $T$. Now the values of
waiting time $t_{\mathrm{w}}$ for each published paper could be
calculated as the number of days between the moment of its
receiving and the closest Editorial Board meeting.

In the absence of peer-review all the received manuscripts are
supposed to be considered by the Editorial Board. In this case the
presence of some constrains is important (for example, the limited
size of the issue). For simplicity all the papers in our model
have an equal number of pages, so the maximal size of a single
issue could be limited by number of papers\footnote{We specify the
issue size equals 10 papers in our models.}. The value of
parameter called the traffic intensity $\rho$ (\ref{traffic_int})
allows us to distinguish three regimes of work
\cite{2006_Vazquez}:
\begin{itemize}
\item the input flow of manuscripts is too slow -- all the papers are accepted for publication but the issue is incomplete (subcritical regime, $\rho < 1$);%
\item the number of received manuscripts is equal to the size of issue -- all the papers are accepted for publication and the issue is complete  (critical regime, $\rho = 1$);%
\item the number of received manuscripts is larger than maximum
number of papers in one issue -- a part of papers are accepted and
the other are waiting in the queue for the next issue
(supercritical regime, $\rho > 1$).%
\end{itemize}

Obviously, the first regime of editorial work is not effective due
to the incompleteness of issues. The third regime is also not
realistic due to the endless queue of papers and therefore
increasingly large values of waiting time. The second regime could
be considered as a perfect one since the issues are complete and
the paper publishing without delays. But this regime is not stable
and it could not be reached in practice because it is impossible
to control the number of input manuscripts from different authors.
It is more probable to provide the critical regime with $\lambda
\approx \mu$, when the queue of papers periodically appears but it
does not grow infinitely. So, the existence of limited queue is
necessary to provide the completeness of issues being some kind of
reserve.

\begin{figure}[!h]
\begin{center}
\includegraphics[width=0.49\textwidth]{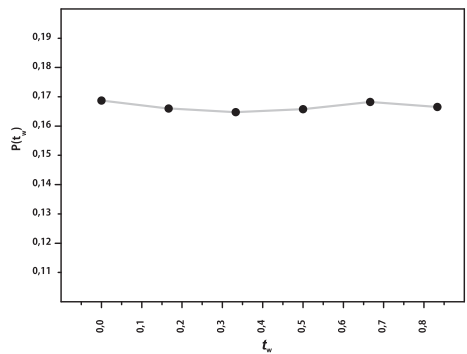}%
\end{center}
\caption{The $P(t_{\mathrm{w}})$ distribution for the modeled
periodical where the number of submitted manuscripts equals to the
number of published papers (determined input flow). }
\label{Fig8_uniform}
\end{figure}

We start modeling with the simplest case when the number of input
manuscripts is determined and equals to the size of issue
(determined input flow).  Besides, we set also the minimal
possible size of printed issued. All the issues which are smaller
than 80\% from standard issue size are considered as incomplete.
As manuscript submission times are uniformly distributed and all
of them are accepted for publication, the resulting
$P(t_{\mathrm{w}})$ distribution is uniform too, see
Fig.~\ref{Fig8_uniform}. In this figure and further each point
means the probability for the published paper to have waiting time
in the interval $t_{\mathrm{w}} \in
[t_{\mathrm{w}}^i..t_{\mathrm{w}}^{i+1})$; the value of
$t_{\mathrm{w}}$ is measured in number of typical journal periods
$T$. The results were obtained from the time-series simulations
each of length 900~$T$.

In practice the number of received manuscripts could not be always
equal to the number of published papers. Traditionally, the input
flow of tasks from an independent sources is modeled as the
Poisson flow:
\begin{equation}
P_k(t)=\frac{\left(\lambda t\right)^k}{k!}\exp(-\lambda t), \qquad
k\geq 0, \qquad t\geq  0, \label{Poisson}
\end{equation}
here, $k$ is the number of input tasks during time interval $t$,
$\lambda$ is the input flow intensity or the number of tasks which
were received during the time unit. Poisson flow is also
characterized by the exponentially distributed time intervals
between two consecutive tasks. So, the modeling of Poisson flow
mostly means generation of series of exponentially distributed
random variables (for example, see \cite{2006_Vazquez}). This
approach is convenient for the cases when the execution law is
also specified by the Poisson law. Then it is possible to study
different work regimes of the system controlling the values of
intensities of input and execution laws.

In the case of our model (omitting the peer-reviewing) we don't
need to specify the execution law since all the papers could be
accepted at one moment at the Editorial Board meeting. So, the
execution rate $\mu$ could be expressed by the number of published
papers while the number of received manuscripts could naturally
define the arrival rate. Consequently, we need to simulate the
input flow in a way allowing to control the input intensity by the
number of  received manuscripts.

We generate the random value distributed by Poisson law
(\ref{Poisson}) with some value of $\lambda$ (input intensity). In
this case different number of manuscripts (which may be larger or
smaller than $\lambda$) could be received during the typical
journal period $T$. Further, having the number of received papers
we distribute them randomly over the $T$.  If the number of
manuscripts exceeds issue size, than a part of them goes to the
queue or could be rejected. There are two ways of manuscript
choosing from the queue (and also from the input flow) in our
model: ``FIFO'' (first-in-first-out) and ``RANDOM''. We can
suppose that the second way could be the simplest method to
reflect the situation with the numerous continuous priorities.

\begin{figure}[!h]
\begin{center}
\includegraphics[width=0.49\textwidth]{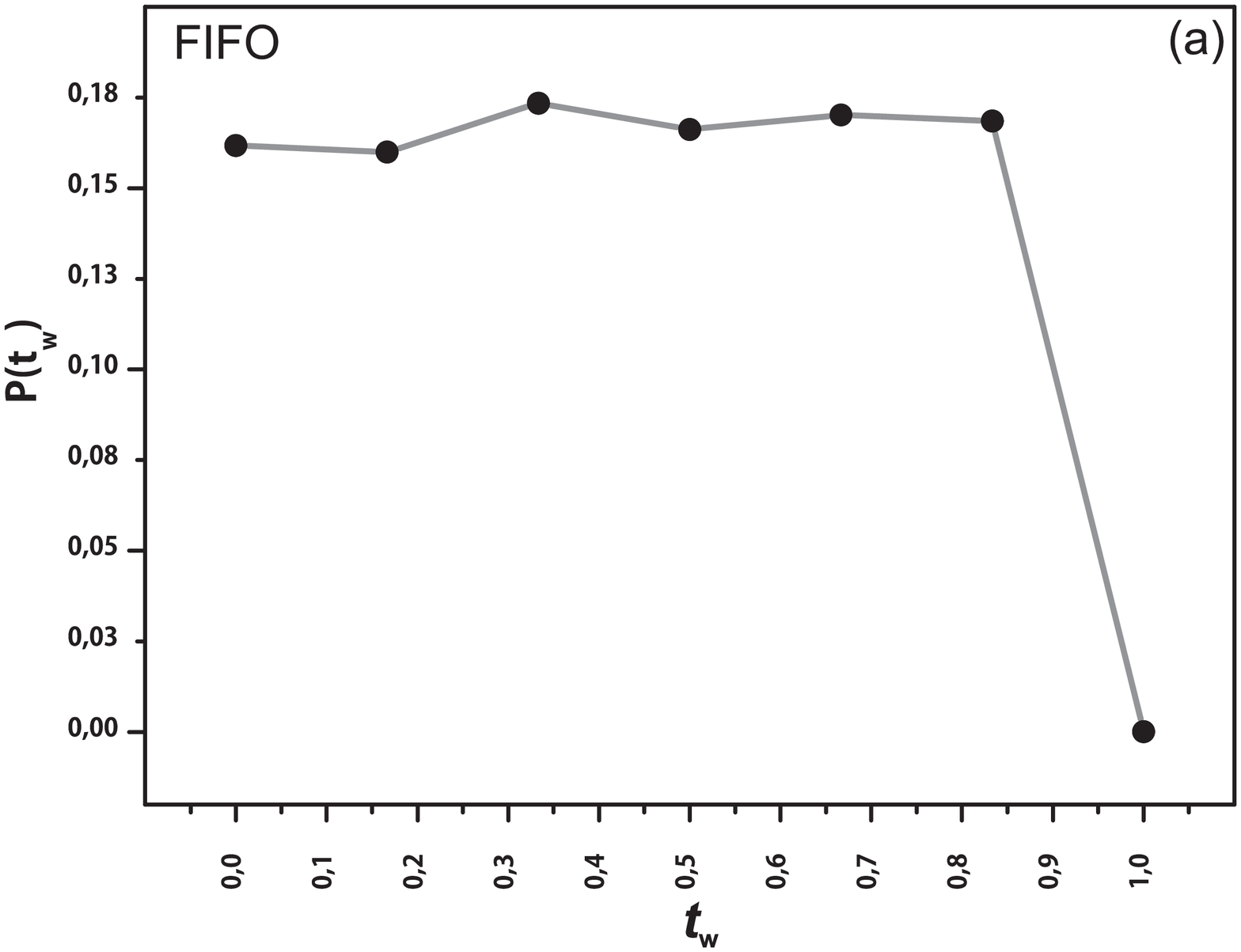}%
\hfill
\includegraphics[width=0.49\textwidth]{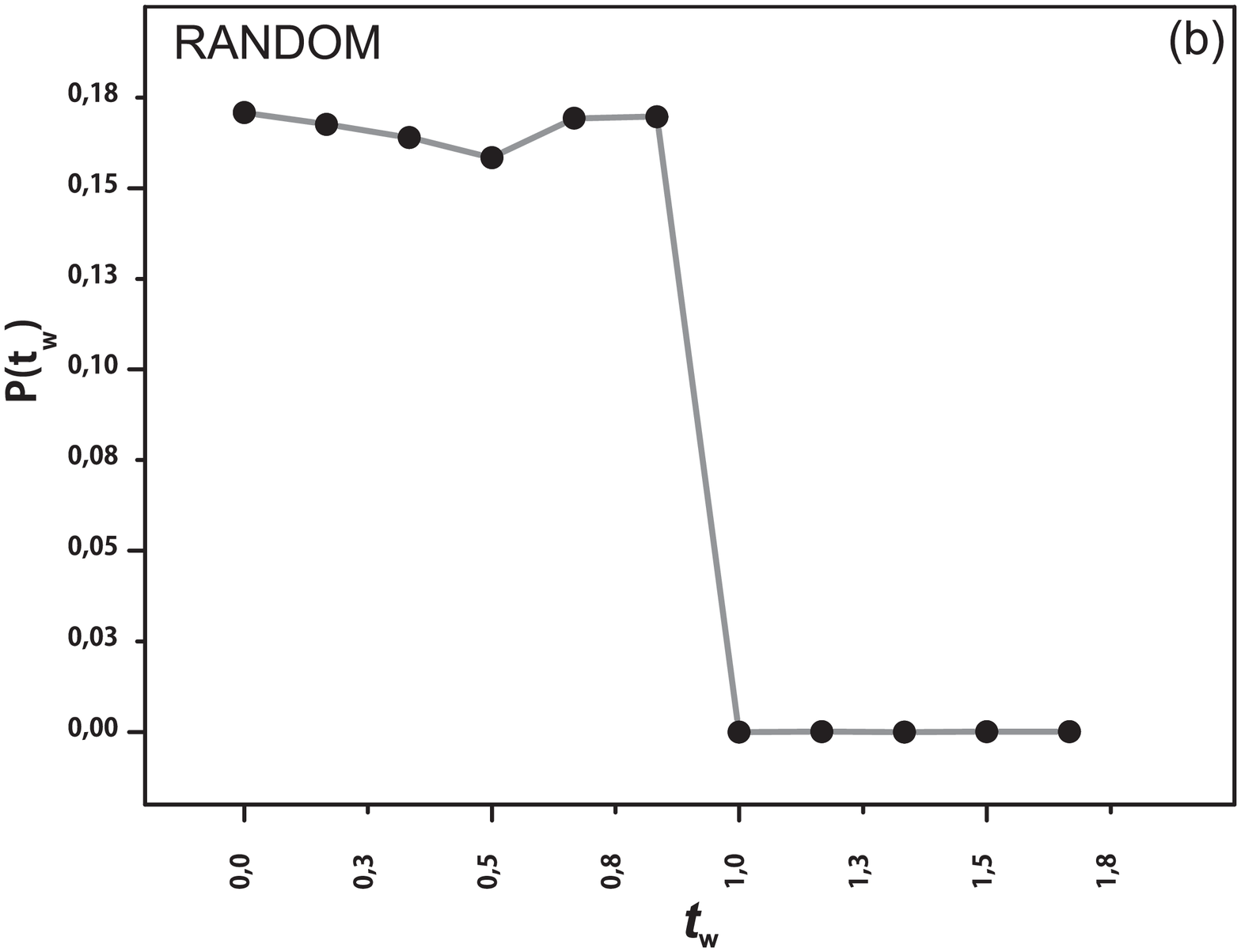}%
\end{center}
\caption{The $P(t_{\mathrm{w}})$ distribution for the modeled
periodical where input rate is comparably small:
$\lambda=3$, $\mu=10$. Two scenarios of manuscript
choosing from queue are used: (a) FIFO and (b) RANDOM.}
\label{Fig9}
\end{figure}
\begin{figure}[!h]
\begin{center}
\includegraphics[width=0.49\textwidth]{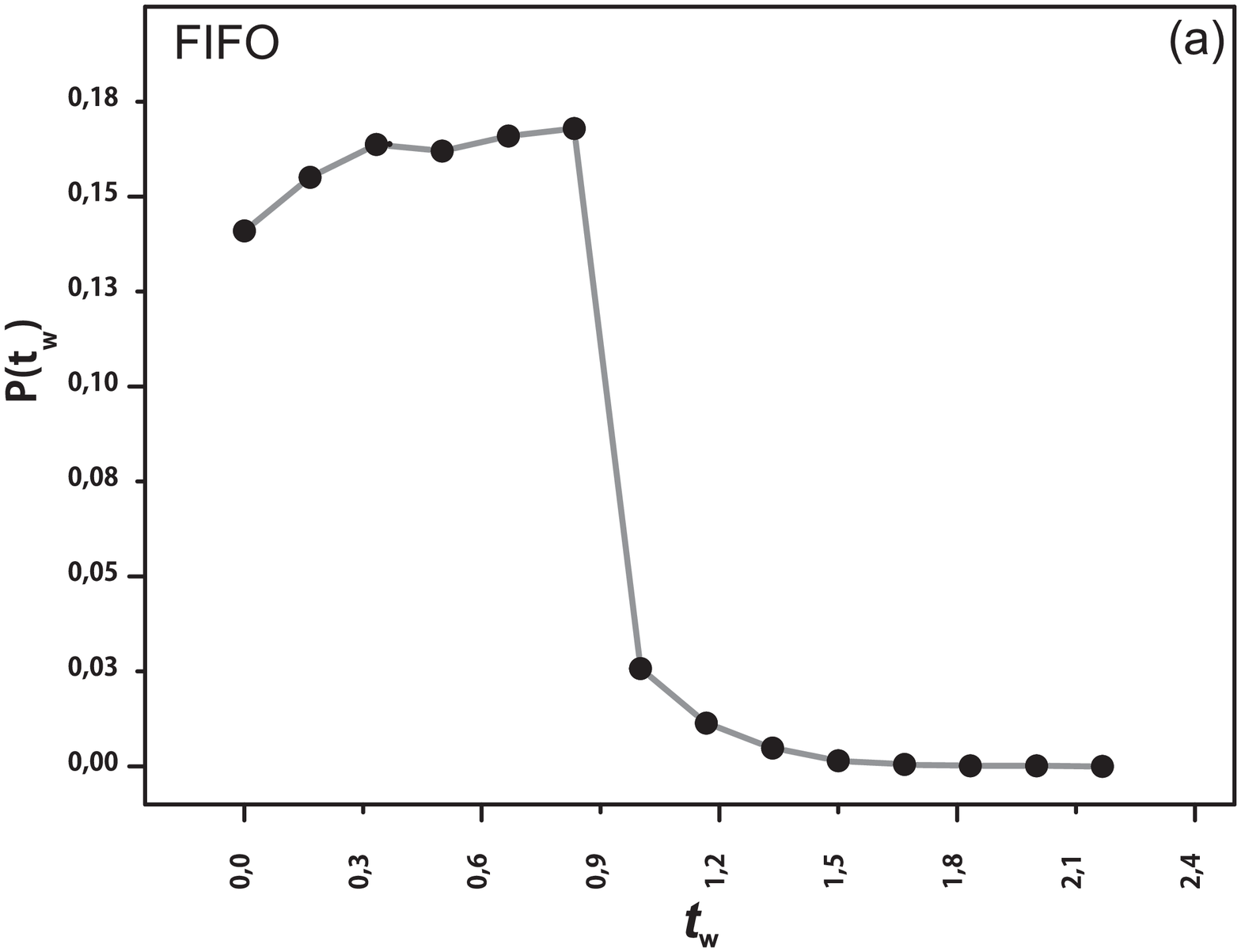}%
\hfill
\includegraphics[width=0.49\textwidth]{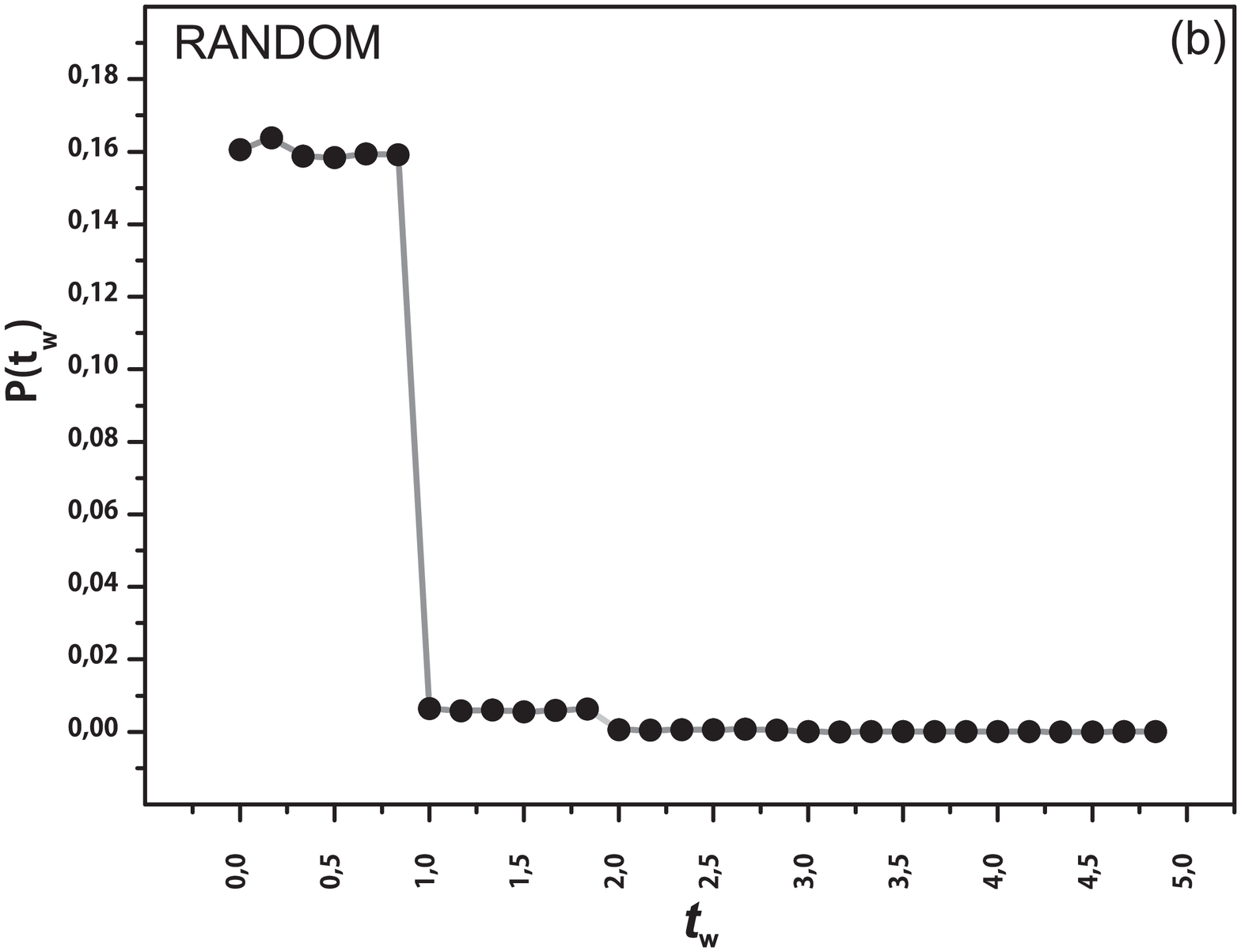}%
\end{center}
\caption{The $P(t_{\mathrm{w}})$ distribution for the modeled
periodical where input rate is approximately equal to the value of
execution rate: $\lambda=7$, $\mu=10$. Two scenarios of manuscript
choosing from queue are used: (a) FIFO and (b) RANDOM.}
\label{Fig10}
\end{figure}

At first we modeled the case with non-determined input flow of
papers and without any limitations of queue length. Modeling the
input flow using (\ref{Poisson}) we can control its intensity
changing the value of $\lambda$. The modeling results are shown in
Figs.~\ref{Fig9} and \ref{Fig10}. Here one can see the change of
$P(t_{\mathrm{w}})$ distribution form according to an increase of
$\lambda$. If $\lambda$ is small then queue doesn't exist and
waiting times are distributed more or less uniformly or forming
the following ``steps''. In this case the majority of published
issues is incomplete (i.e., above 98\% incomplete issues in the
case of $\lambda=3$ and $\mu=10$).

\looseness=-1The queue appears when $\lambda$ is of the order of
magnitude of $\mu$. In this (critical) regime the queue length
fluctuates around some value: the ``excess'' manuscripts from
previous periods can fill the deficiency of papers for next
periods. The $P(t_{\mathrm{w}})$ distribution in this case has two
distinct ``steps'' (Fig.~\ref{Fig10}) presenting two categories of
papers: which are printed in the first following issue and which
are reserved for next periods in the queue. In our model such
critical regime was reached at $\lambda=7$ and $\mu=10$. Then the
number of incomplete issues descends down to 56\%.

\begin{figure}[!h]
\begin{center}
\includegraphics[width=0.49\textwidth]{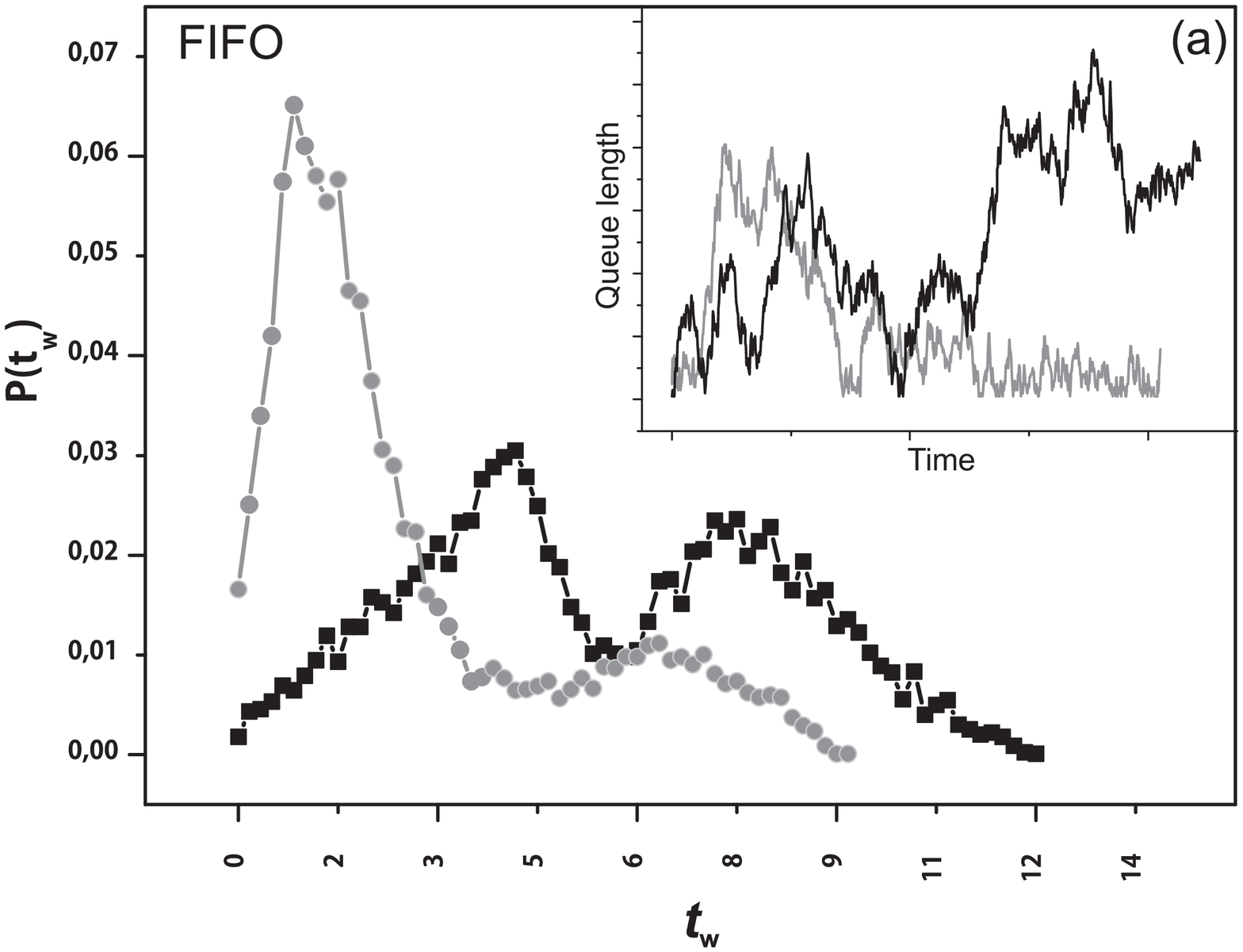}%
\hfill
\includegraphics[width=0.49\textwidth]{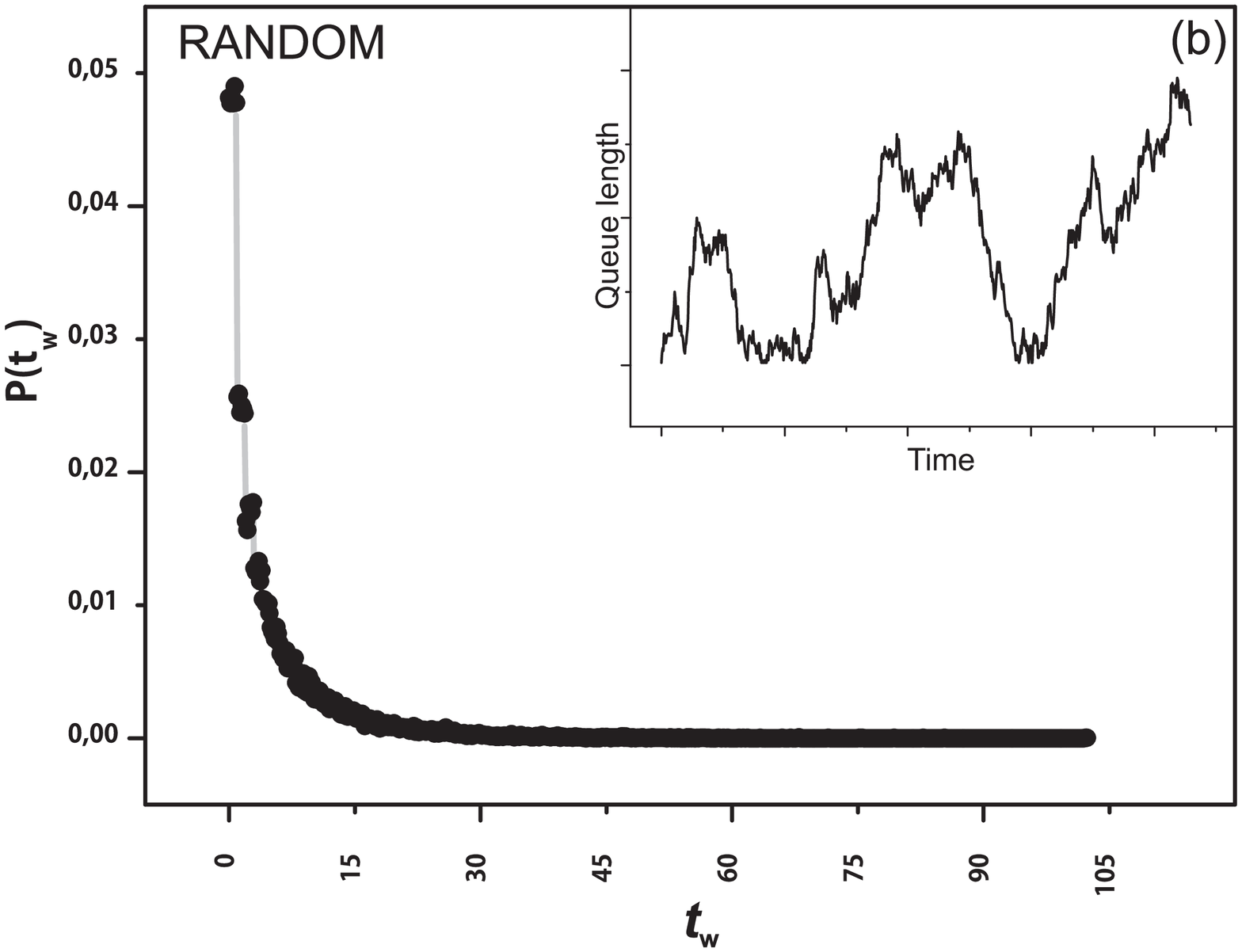}%
\end{center}
\caption{The $P(t_{\mathrm{w}})$ distribution for the modeled
periodical with $\lambda=10$ and $\mu=10$. Two scenarios of
manuscript choosing from queue are used: (a) FIFO (two cases) and
(b) RANDOM. The queue length growing is shown on the corresponding
insets.}
\label{Fig11}
\end{figure}

To minimize part of incomplete issues it is enough to increase
$\lambda$ still more. Our modeling results at $\lambda=10$ and
$\mu=10$ are presented in Fig.~\ref{Fig11}. This case corresponds
to the unsteady state of the system. Despite the fact that number
of incomplete issues is close to zero, such work regime is not
efficient due to increasing queue length (one can see the queue
length growing on the corresponding insets in Fig.~\ref{Fig11}).

\begin{figure}[!h]
\begin{center}
\includegraphics[width=0.49\textwidth]{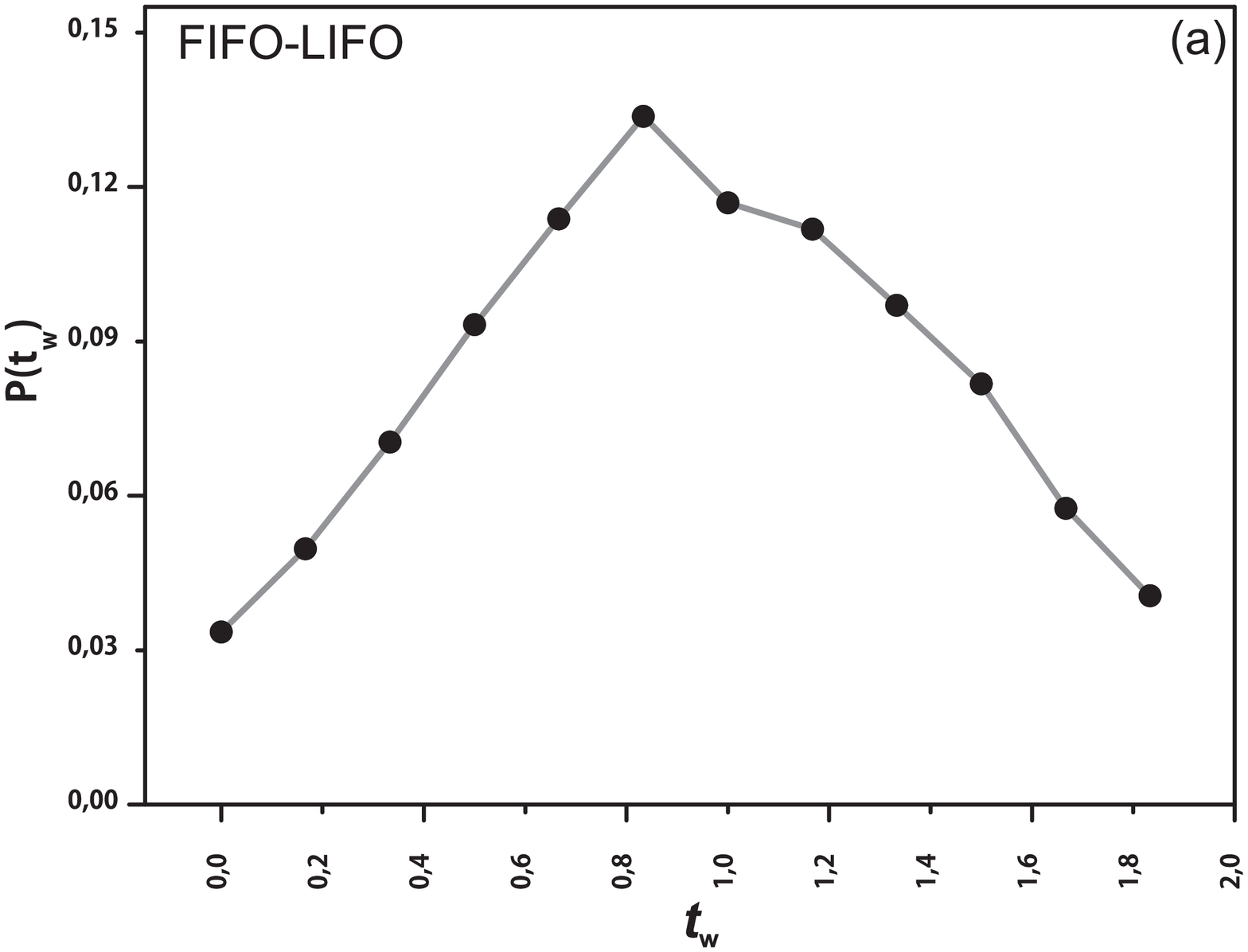}%
\hfill
\includegraphics[width=0.49\textwidth]{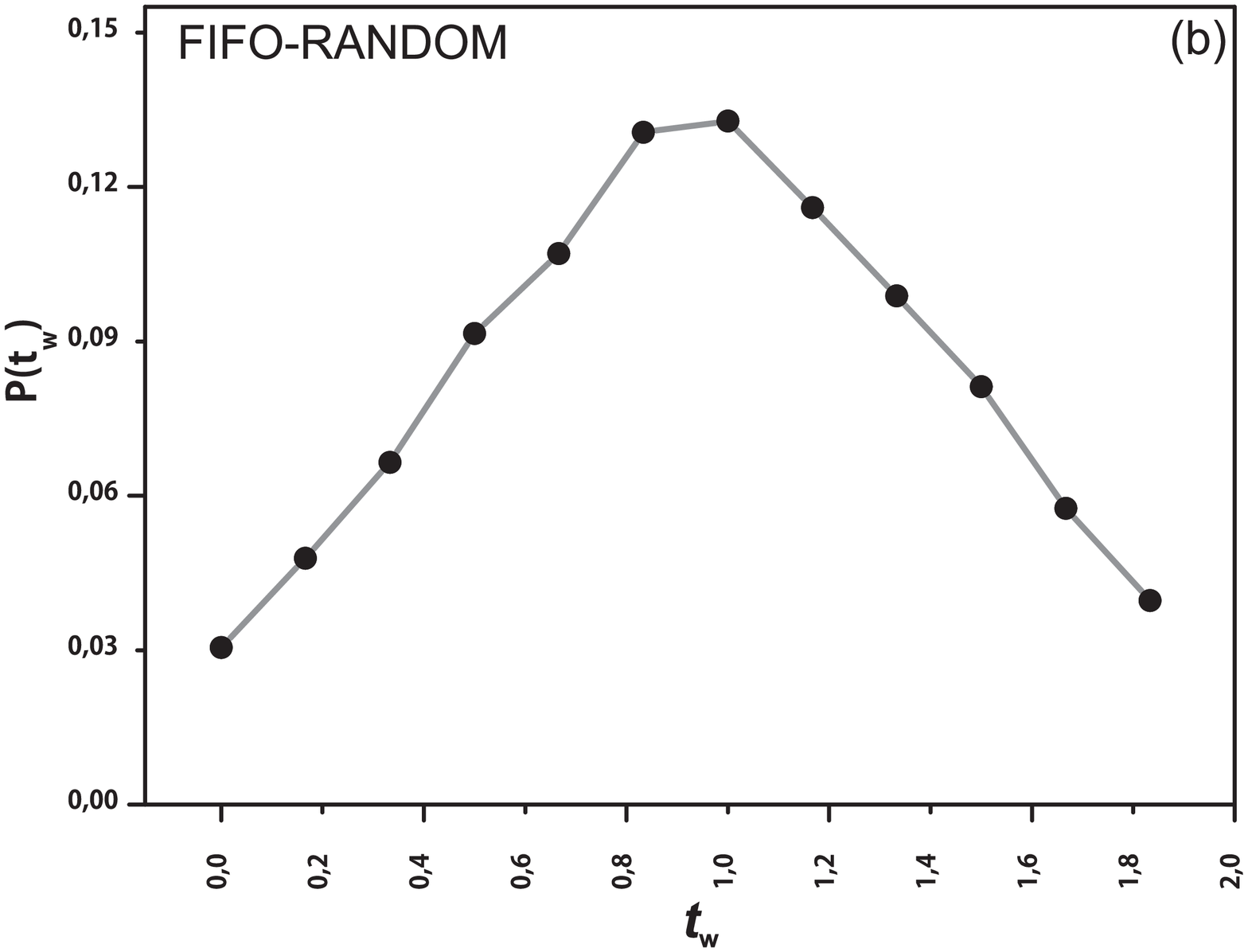}%
\\
\includegraphics[width=0.49\textwidth]{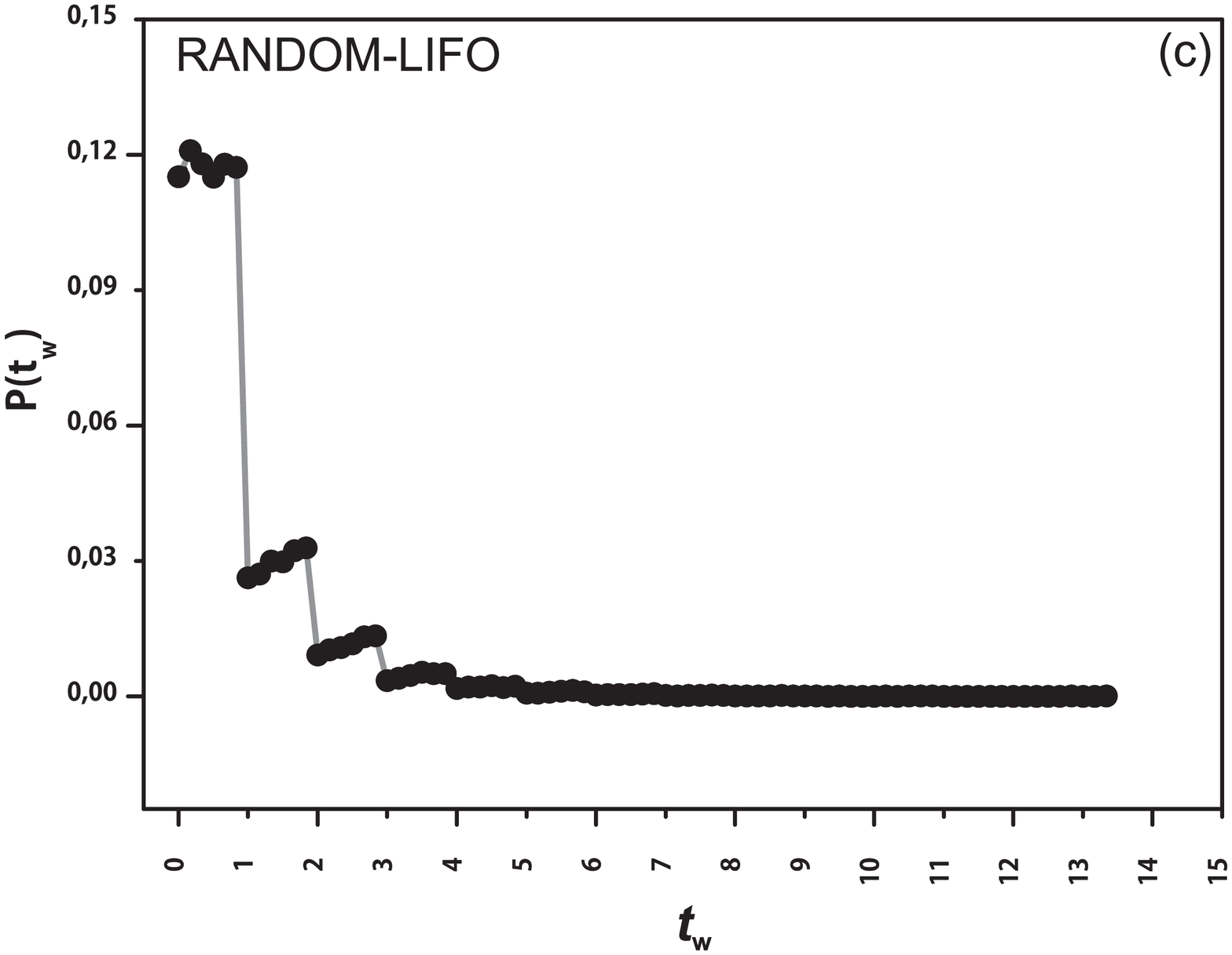}%
\hfill
\includegraphics[width=0.49\textwidth]{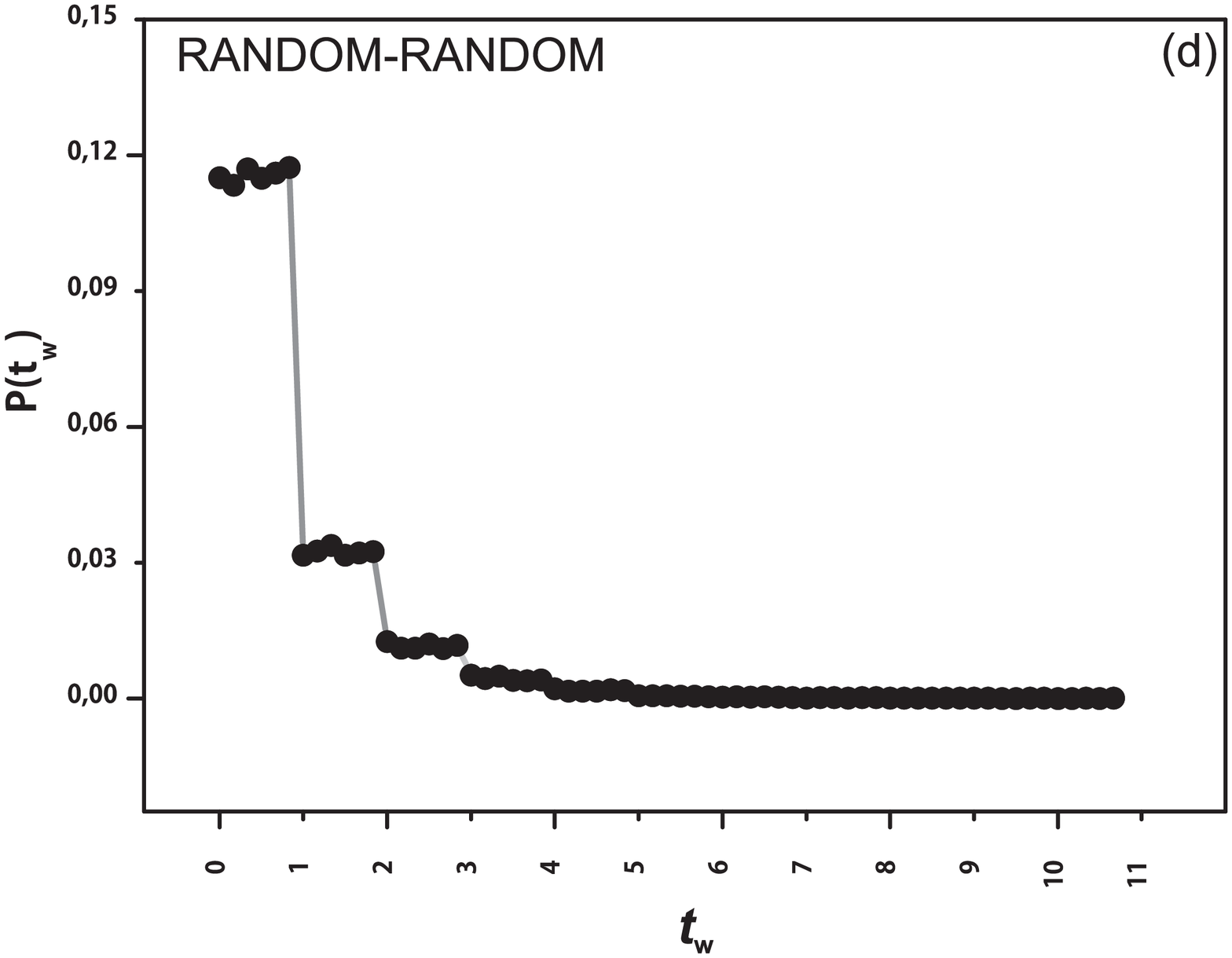}%
\end{center}
\caption{The $P(t_{\mathrm{w}})$ distribution for the modeled
periodical with $\lambda=10$ and $\mu=10$ using (a), (b) FIFO and
(c), (d) RANDOM manuscripts selection rules. Two scenarios of
excess papers rejection are used: (a), (c) LIFO and (b), (d)
RANDOM.}
\label{Fig12}
\end{figure}
We can conclude that it is impossible to reach the optimal work
regime for Editorial Board in scientific journal without any
artificial constrains. So, the next steps in description of this
process may involve additional constrains into the model.
 We limit
the queue length to the issue size specified before. In our case
normal issue volume equals 10, so the queue length is limited to
10 also. Thus, such work regime could be called ``the issue in
reserve''. According to our rules the excess papers should be
rejected using one of the scenarios: ``LIFO'' (last-in-first-out)
and ``RANDOM''.

As we can see from Fig.~\ref{Fig12}, the obtained functional form
of $P(t_{\mathrm{w}})$ distribution considerably differs from the
experimental one. When the ``FIFO'' scenario is applied to select
the manuscript from a queue the waiting time distribution is
symmetric with one distinct maximum (Fig.~\ref{Fig12} (a) and
(b)). In the case of ``RANDOM'' selection rule the waiting time
distribution has decreasing form but also several ``steps''
corresponding to each time period could be observed
(Fig.~\ref{Fig12} (c) and (d)). Besides, the rate of decreasing
observed is larger (with value of exponent close to 3) and more
close to the exponential function.

To summarize this section several aspects should be mentioned.
First of all, it is important to note, that it is impossible to
any real scientific journal to publish all the manuscripts
received. So, every Editorial Board should choose the way or rule
to eliminate the part of them: examining the execution of some
obligatory technical requirements, applying peer-review mechanism,
or following by other own internal criteria. The second important
remark is that one can suppose that for different scenario it is
possible to notice different picture of waiting time distribution
for the published manuscripts. In this section we made the attempt
to simulate most simple case of manuscripts arrangement in
scientific journal, i.e. without any external peer review but
using just simple rules. The obtained functional form of papers
waiting time distributions for this scenario differs from the
analogous results obtained for the real journals with peer-review
mechanism. Of course, it is also interesting to simulate exactly
the work with peer-reviewing but the way of natural human activity
simulation is the subject of the discussions yet \cite{2008_Zhou}.

\section*{Conclusions}
We have found that both log-normal and power-law function with
exponential cutoff and exponent $\alpha=1$ can be the probable
functions of waiting time distributions $P(t_{\mathrm{w}})$ for
manuscripts in scientific journals. In fact, both log-normal and
power-law functions predict exactly the same leading power
behavior $t^{-1}$, differing only in the functional form of the
exponential correction \cite{2005_Barabasi}. In this sense,
process considered here is governed by similar probability
distributions as other examples of human activities
\cite{2005_Barabasi_Nature,2005_Oliveira&Barabasi_Nature,2008_Zhou,2004_Johansen,2006_Vazquez,2006_Stouffer}.
The observed data fluctuations can be explained by relatively
small statistics but such situation is usual for the majority of
real-world data bases. Thus, we consider the obtained form of
probability distributions $P(t_{\mathrm{w}})$ as the typical one
that can be used for scientometrical analysis of editions. The
length of waiting times is an important characteristic of
Editorial Board's activity. The publication delay effects journal
rankings according to the impact factor \cite{2006_publ_delay} as
well as personal citation rating of authors.

The simple model of Editorial Board work was created to verify the
hypothesis about the important role of peer-reviewing in the
waiting times distributions $P(t_{\mathrm{w}})$ shaping. In
general, the obtained results can be considered as the support of
our previous conclusions.

In conclusion it is worth to mention some peculiarities of
scientific editorial process that were not taken into
consideration in our work. Here the main idea is connected with
emphasizing of the meaning of natural human activity within the
complex processes taking place in the Editorial Boards working.
But these natural human activity patterns could be also affected
by different situations or rules. For example, journal policy
could be very different concerning the strictness of time
deadlines for peer-reviewing or concerning the manuscripts with
one ``positive'' and one ``negative'' referee reports.
In addition to peer review we can mention other processes as the
examples of natural human activity: technical work with raw
manuscripts, communication between the editors and other
participants of the editorial processes, etc. Moreover, one can
distinguish several human activity sub-processes within peer
reviewing such as pure referee work with manuscript, communication
processes between referee, editorial office and authors, and,
eventually, the manuscript revision by authors according to the
referee remarks. It is very complex problem to account all these
aspects in one model. So, it could be the worth challenge for the
future analysis.

\section*{Acknowledgements}
This work was supported in part by the bilateral cooperation
project ``Scientometrics: quantitative approach to social
phenomena'' (SCSII, Ukraine and the Ministry of Research and
Science, Austria) and the 7th FP, IRSES project N269139 "Dynamics
and Cooperative phenomena in complex physical and biological
environments". The authors thank Dr. R. Kenna (Coventry) for
useful suggestions and comments.


\end{document}